\documentclass[11pt,preprint2]{aastex}

\usepackage {pstricks}

\shorttitle{Turbulent Cells}
\shortauthors{Arnett \& Meakin}

\def \eavg#1{\relax \langle #1 \rangle}

\newcommand{\etal} { et~al.\ }  

\newcommand{\mconv}{{\cal M}_{conv}}

\def \nuc#1#2{\relax\ifmmode{}^{#1}{\protect\text{#2}}\else${}^{#1}$#2\fi}
\def \msol#1{\relax$#1\,M_\odot\/$}

\begin{document}

\title{Turbulent Cells in Stars: Fluctuations in Kinetic Energy and Luminosity}

\author{W. David Arnett\altaffilmark{1,2}
\altaffiltext{1}{Steward Observatory, University of Arizona, 
933 N. Cherry Avenue, Tucson AZ 85721}
\altaffiltext{2}{ICRAnet, Rome, Pescara, Nice}
\email{ wdarnett@gmail.com}
}
\author{Casey Meakin\altaffilmark{1,3}
\email{casey.meakin@gmail.com }
\altaffiltext{3}{Theoretical Division, Los Alamos National Loboratory,
Los Alamos, NM 875, USA}
}

\begin{abstract}
Three-dimensional (3D) hydrodynamic simulations of shell oxygen burning \citep{ma07b} 
exhibit bursty, recurrent fluctuations in turbulent kinetic energy. These are shown to be due to
a general instability of the convective cell, requiring only a localized source of heating  or cooling.  Such fluctuations are shown to be suppressed in 
simulations of stellar evolution which use mixing-length theory (MLT). 

Quantitatively similar behavior occurs in the model of a convective roll (cell)
 of \cite{lorenz}, which is known to have a strange attractor that gives rise to 
 chaotic fluctuations in time of velocity and, as we show, luminosity. 
 Study of simulations suggests that the behavior of a Lorenz convective roll may resemble that of  a cell in convective flow. 
 We examine some implications of this
 simplest approximation, and suggest paths for improvement.  

Using the Lorenz model as representative of a convective cell,
a multiple-cell model of a convective layer gives total  luminosity fluctuations
which are suggestive of irregular variables (red giants and supergiants
\citep{ms75}), and of the long secondary period feature in semiregular AGB
variables \citep{stothers,wok04}.  This ``$\tau$-mechanism" is a new source for stellar variability, which is inherently non-linear (unseen in linear stability
analysis), and one closely related to intermittency in turbulence. 
It was already implicit in the 3D global simulations of \cite{wpj03}.
This fluctuating behavior is seen in extended 2D simulations of CNeOSi burning
shells \citep{am11b}, and may cause instability
which leads to eruptions in progenitors of core collapse supernovae {\em prior} to
collapse.

\end{abstract}


\section{Introduction}

Three-dimensional fluid dynamic simulations of turbulent convection in an
oxygen-burning shell of a presupernova star show bursty fluctuations which
are not seen in one-dimensional stellar evolutionary calculations (which use
various versions of mixing-length theory, MLT, \cite{bv58}). This paper
explores the underlying physics of this new phenomena.

Since the formulation of MLT \citep{bv58}, 
there have been a number of significant developments in the theoretical
understanding of turbulent convective flow.

First, \cite{kolmg} and \cite{obukhov} developed the modern version of the
turbulent cascade, and published in journals easily accessible in the
west; the original theory \citep{kolmg41} was not used 
in MLT although it pre-dated it. This explicit expression for dissipation
of turbulent velocities\footnote{This is essentially the ``four-fifths law of Kolmogorov", \cite{frisch}, p. 76.},
\begin{equation}
\epsilon_{turb}  =  u_{rms}^3 /\ell_{d} ,
\end{equation}
where $u_{rms}$ is the root-mean-square of the turbulent velocity and
$\ell_{d}$ is the dissipation length. It is found both experimentally and
numerically that $\ell_{d} \approx \ell_{CZ}$, where $\ell_{CZ}$ is the depth
of the convective zone. Simulations for low-Mach number flow show that
the average of this dissipation over the convective zone closely compensates
for the corresponding average of the buoyant power \citep{amy09}.
This additional constraint allows an alternative to present practice: fixing the free parameter (e.g., the mixing length factor $\alpha$) directly by terrestrial experiments and numerical simulations which deal with the process of turbulence itself \citep{amy10}, instead of calibrating it from complex astronomical systems (stellar atmospheres) as is now done.

Second, there has been a considerable development
in understanding the nature of chaotic behavior in nonlinear systems;
see \cite{cvit} for a review and reprints of original papers,
and \cite{frisch,gleick,thomp}. 
\cite{lorenz} presented a simplified solution to the Rayleigh problem
of thermal convection \citep{ch61} which captured the seed of
chaos in the Lorenz attractor, and contains a representation of the
fluctuating aspect of turbulence not present in MLT. This advance was
allowed by the steady increase in computer power and accessibility,
which lead to the exploration of solutions for simple systems of nonlinear
differential equations (see \cite{cvit} and references therein).
It became clear that the Landau picture \citep{landau44} of the approach to turbulence
was incorrect both theoretically \citep{rt71} and experimentally \citep{libchaber}.
A striking feature of these advances has been the use of simple mathematical
models, which capture the essence of chaos in a model with much reduced
dimensionality compared to the physical system of interest.

Third, it has become possible to simulate turbulence on computers.
This realizes the vision of John von Neumann \citep{vonneuman48},
in which numerical solutions of the Navier-Stokes equations by computer
are used to inform mathematical analysis of turbulence. In this paper we
will follow this idea of von Neumann, in the style which proved successful for chaos
studies: building simple mathematical models of a more complex physical
system (in this case, the numerical simulations of turbulent convection). 
This approach should lead to algorithms suitable for implementation into
stellar evolution codes, which, unlike MLT, are (1) based upon solutions to 
fluid dynamics equations, (2) non-local, (3) time-dependent, and (4) 
falsifiable by terrestrial experiment and future improved simulations.

Our particular example is a set of simulations of oxygen burning in a shell of
a star of $23 \rm M_\odot$ \citep{ma07b}. This is of astronomical interest 
in its own right as a model for a supernova progenitor, but
also happens to represent a relatively simple and computationally efficient case,
and has general implications for the convection process in all stars.

Three-dimensional hydrodynamic simulations of shell oxygen burning
exhibit bursty, recurrent fluctuations in turbulent kinetic energy (\cite{ma07b} and below).
The reason for this behavior has not been explained theoretically.
These simulations show a damping, and eventual cessation, of turbulent motion 
if we artificially turn off the nuclear burning \citep{amy09}. Further investigation \citep{ma09} shows that nearly identical pulsations are obtained with a volumetric energy generation rate which is constant in time, so that {\em the cause of the pulsation is independent of any temperature or composition dependence in the oxygen burning rate.} Localized heating is necessary to drive the convection; even with this  time-independent rate of heating,  pulses in the turbulent kinetic energy still occur.

Such behavior is fundamentally different from traditional nuclear-energized 
pulsations dealt with in the literature (e.g., the $\varepsilon$-mechanism, \cite{ledoux41,unno89}),
and is a consequence of time-dependent turbulent convection (it might be called
a "$\tau$-mechanism", with $\tau$ standing for turbulence).
It appears to be relevant to all stellar convection. \cite{wpj03} found, in a very different context, that non-linear interaction of the largest modes excited pulsations of a red-giant envelope\footnote{A $\kappa$-mechanism, which depends upon variations in opacity, is not required to drive such pulsations.}, which is another example of the $\tau$-mechanism.

In Section~2 we examine the the physics context of the turbulence,
including implications of subgrid and turbulent dissipation
for the implicit large eddy simulations (ILES) upon which our analysis is based,
and the effect of the convective Mach number on the nature of the flow.
In Section~3 we review the 3D numerical results of shell oxygen burning 
which are relevant to the theory.
In Section~4 we present the results of the classical Lorenz model \citep{lorenz} 
for conditions similar to those in Section~3.
In Section~5 we consider implications of turbulent intermittency on stellar
variability, and provide a model light curve from this effect alone. 
Section~6 summarizes the results. The appendix gives a short derivation of the Lorenz model.

\section{Physics Context.}
In this section we summarize concepts which are needed for the
interpretation of later results.

\subsection{Subgrid Damping and Kolmogorov\label{ILES}}

Approximation of partial differential equations by discrete
methods inevitably leads to a loss of information at scales smaller
than the grid size. A single element in space is approximated
as a homogeneous entity\footnote{Actually most modern simulations (ours included) use higher order methods which make some further assumptions regarding the behavior of variables inside a zone. This complicates but does not change the argument; it is still true that information is lost at the zone level.}; 
this is equivalent to complete mixing at this
scale, at each time step, of mass, momentum, and energy. The loss
of information that occurs with this mixing corresponds to an 
increase in entropy \citep{shannon}, the mixing of  momentum is 
equivalent to the action of viscosity,  and the mixing of internal energy
corresponds to the transport of heat (\cite{ll59}, \S15 and \S49). 

In 3D flow, turbulent energy will cascade from large
scales to small, at a rate set by the largest scales \citep{kolmg41}. At sufficiently 
small scales, microscopic processes homogenize the flow and dissipate
the kinetic energy. Thus, there is a deep connection between the 
turbulent cascade and sub-grid scale mixing.

\cite{sytine} have demonstrated
that the piece-wise parabolic method (PPM, which we use), based on the Euler equation
(which has no explicit viscosity), converges to the same limit as methods based on 
Navier-Stokes equation  (which do have explicit viscosity),
as the grid is refined to smaller zones and smaller effective viscosity 
(the relevant limit for astrophysics).  The subgrid scale dissipation
for monotonicity preserving hydrodynamic algorithms \citep{boris,woodward}, 
which is implicit in these methods,
automatically gives a reasonable treatment of the turbulent cascade down
to the grid scale.
We use this implicit sub-grid dissipation in our large eddy 
simulation (ILES); this is the most computationally
efficient way to deal with turbulent systems with a large range of scales.
The largest scales, which set the rate of cascade and
contain most of the energy are explicitly calculated, while the 
sub-grid scales are dissipated in a way consistent with the Kolmogorov
cascade.

\cite{wood06} have presented a refinement of the PPM algorithm which has
improved behavior at the smaller resolvable scales. We have not yet implemented this modification. The theoretical approach
used here involves integrated properties of convective cells; we find
by direct resolution studies that these properties are well estimated even with
surprisingly modest resolution, because they are determined primarily by the largest
scales in the convective region. It appears that the ILES simulations are adequate 
for the present analysis. 

\cite{amy09} have shown that the numerical damping at sub-grid scales 
in our ILES simulations is quantitatively consistent with the introduction {\em analytically}  of the Kolmogorov cascade into the theoretical discussion. 
The turbulent velocity field was found to be dominated by two components:
\begin{enumerate}
\item a non-isotropic flow of the largest scale modes in the convection zone, 
which is coupled to the fluctuations. This has aspects of a ``coherent structure''
\citep{hlb96}.
The largest scales are unstable toward break-up, but are least affected by
dissipation, and in this sense the most laminar.
\item a more isotropic, homogeneous turbulent flow
which carries the kinetic energy via the turbulent cascade to scales small
enough for dissipation to occur \citep{kolmg}.
Because of the vast size difference, the small scales are weakly coupled to
the largest scales, which determine the rate at which energy flows through
the cascade. 
\end{enumerate}
If  we approximate the non-isotropic component of the flow (the largest scale of convection) with that described by the Lorenz model,
this interpretation captures the oxygen-burning fluctuations.

\subsection{Types of Flow\label{aake}}
There are two limiting cases for convective flow, depending upon the convective
Mach number $\mconv$ (the ratio of the fluid speed to the local sound 
speed); these are usually termed the ``incompressible"  ($\mconv \ll 1$)
and ``compressible" ($\mconv \sim 1$) regimes. Stars are stratified in density, so that the notion of ``incompressiblity" is
misleading. We will use the term ``low-Mach number flow" 
in place of ``incompressible" when we mean flows in which acoustic radiation
is small, but may be compressed due to stratification.

For turbulent motion, the pressure perturbation $P'$ is related to the convective Mach number by $P'/P \sim \rho u_{rms}^2/P  \sim \mconv^2$.
Sound waves outstrip fluid motion, so that pressure differences quickly become small, except
possibly for a static background stratification. Most of the historical research on convection
(e.g., the B\`enard problem, \cite{ch61,ll59}) is done in this limit.

Using the Reynolds decomposition, $\varphi = \varphi_0 + \varphi'$, with
horizontal averaging $\langle \varphi \rangle = \varphi_0$ and
$\langle \varphi' \rangle = 0$,
mass conservation for a steady state flow can be written 
\begin{equation}
 \langle \rho{\bf u} \rangle=  \langle \rho_0 {\bf u_0}\rangle + 
 \langle \rho' {\bf u'}\rangle = 0. 
\end{equation}

The Navier-Stokes equation,  using mass conservation, is
\begin{equation}
\partial_t \rho{\bf u}+ {\bf \nabla \cdot} \rho {\bf u u} = - \nabla P
+ \rho  {\bf g} + \nu \nabla^2 {\bf u},
\end{equation}
where $\rho {\bf uu}$ is the Reynolds stress tensor.
Mass conservation implies a convenient identity involving total co-moving
derivatives,
\begin{equation}
D_t \rho {\bf u}  = \rho D_t {\bf u}.
\end{equation}
Taking the dot product of the velocity vector $\bf u$ with the Navier-Stokes equation, gives a kinetic energy equation,
\begin{equation}
D_t \rho{\bf u\cdot u}/2 = -{\bf u \cdot \nabla} P
+ \rho {\bf u \cdot g} + \nu {\bf u \cdot}\nabla^2 {\bf u}. \label{ke-eq}
\end{equation}
If on average the system is in a steady state, the time derivative must integrate to
zero over the convective region, and the mass conservation law implies that the total buoyancy power term is zero, $\eavg{\rho {\bf u \cdot g}} =0$ (assuming constant $\bf g $ on horizontal averaging), and therefore does not contribute to the production of kinetic energy anywhere in the flow.  The only other term, which remains to balance the viscous dissipation of the kinetic energy, is the pressure term 
\begin{equation}
{\bf u \cdot \nabla}P= {\bf u_0 \cdot \nabla} P_0 + \eavg{ {\bf u' \cdot \nabla} P' }.
\end{equation}
which may be rewritten as
\begin{equation}
{\bf u \cdot \nabla}P= {\bf u_0 \cdot \nabla} P_0 + \eavg{ \nabla \cdot (P' {\bf u'}) }
- \eavg{ P' {\bf \nabla \cdot u' }}.
\end{equation}
The divergence term vanishes upon integration over the volume.
Using hydrostatic equilibrium for the background state,  ${\bf \nabla} P_0 =  \rho_0 {\bf g}$,  and mass
conservation, $ \rho_0 {\bf u_0}  = -\langle \rho' {\bf u'} \rangle,$
\begin{equation}
\langle -{\bf u \cdot \nabla} P \rangle =  \langle \rho' {\bf u' \cdot g} \rangle
+ \langle P' {\bf \nabla \cdot u'} \rangle \label{pfluct}
\end{equation}
When the Mach number is small, the second term the right hand expression is nearly zero because $\bf \nabla \cdot u' \approx 0$ and the turbulent pressure is negligible. 
In this limit the kinetic energy production is best understood as due to the remaining buoyancy power term $\eavg{\rho' {\bf u' \cdot g}}$, which is directly related
to the enthalpy flux \citep{amy09}.  

When the Mach number is no longer small, the second term on the RHS of
Eq.~\ref{pfluct} increases in importance: both the divergence of the fluctuating velocity field and the pressure perturbation begin to play a role.  The velocity field
changes character; it is no longer dominated by  rotational flow, but develops an irrotational component (${\bf \nabla \cdot u'} \neq 0)$. The flow becomes diverging  (consider the extreme limit of a point explosion which is pure divergence). 
Also, the ram ``pressure" (a tensor $ \rho {\bf u u}$) is not negligible and must be included in the momentum equation (``hydrostatic equilibrium").
Sound wave generation increases rapidly as  $\mconv \rightarrow 1$ (\cite{ll59}, \S75). 
The compressible limit is $\mconv \simeq 1$. Shock formation is the most startling change in the flow character. 

Which of the $\mconv$ limits is relevant for astrophysics? Both are. 
Almost all the matter in stellar
convection zones, during almost all evolution, is in the limit of low-Mach number flow, as are our turbulence simulations. The exceptions are important: (1) explosions, such as supernovae and novae, (2) vigorous thermonuclear flashes, (3) vigorous pulsations, especially radial ones, and (4) the sub-photospheric layers of stellar surface convection zones, which are strongly non-adiabatic, to name a few. 

\section{The Oxygen Shell Simulation}

\begin{figure}
\figurenum{1}
\includegraphics[totalheight=4.2in,origin=c]{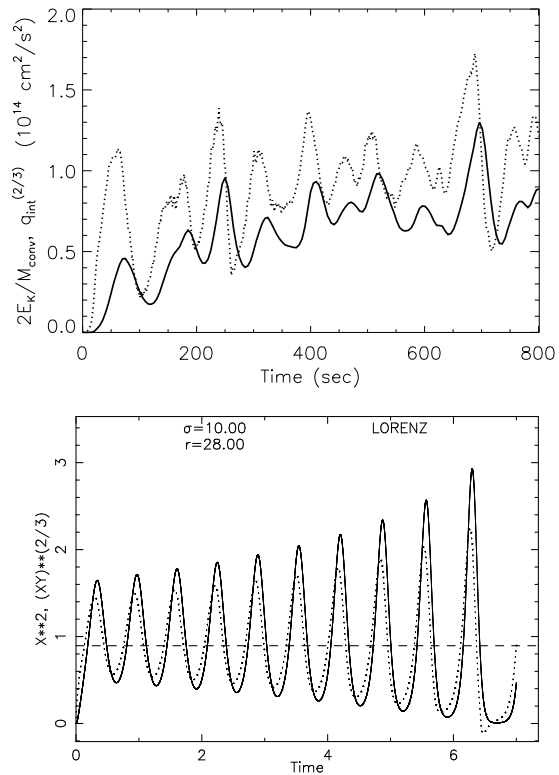}
\caption{Comparison of Convective Kinetic Energy Fluctuations in the
3D simulations and in the Lorenz model, starting from zero velocity. 
The Lorenz model is labeled with its Prandtl number $\sigma$ and its Rayleigh
number $r$.
Turbulent velocity squared (solid) and buoyancy power to the 2/3 power (dotted) are plotted versus time. The axes where chosen so that the same
 number of peaks would be shown, and the average kinetic energy be the same. 
The curves are surprisingly similar, even though (1) the Lorenz model used only a single mode, (2) $\sigma$ and $r$ were not adjusted to give a fit,
(3) the Lorenz model imposes thermal balance while thermal imbalance adds to
the background slope in the simulation (see Fig.~\ref{fig2}),  and (4) the simulations have $\sim 10^8$ more degrees of freedom than the
Lorenz model.
}
\label{fig1}
\end{figure}

\begin{figure}
\figurenum{2}
\includegraphics[angle=-90,scale=0.3]{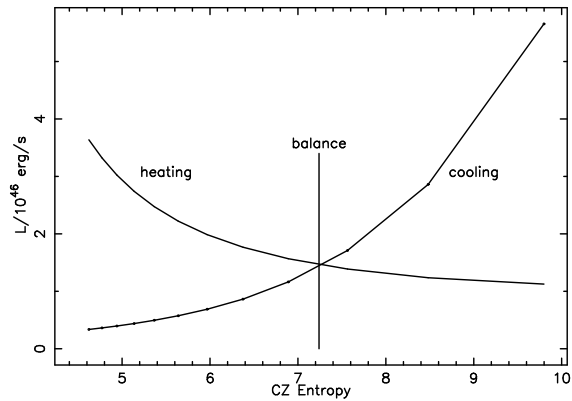}
\caption{Nuclear heating and neutrino cooling
in an oxygen burning shell, as a function of entropy. At lower entropy,
as in the simulations,
nuclear heating dominates because of its density dependence.
For the same temperatures but higher entropy, the inhibition of
neutrino-antineutrino emission is reduced. The shell entropy slowly
increases until cooling balances heating, at which point 
thermal balance is attained.
}
\label{fig2}
\end{figure}
\placefigure{2}

Figure~\ref{fig1} illustrates the behavior of two important integral quantities, 
the total turbulent kinetic energy and the total buoyancy power,
in the oxygen-burning shell simulations for a 23 $\msol$ star 
\citep{ma07a}. The flow has a low-Mach number ($\mconv^2 \le 0.001$),
although the numerical simulations use the equations for fully compressible
fluid flow, and would have correctly treated high-Mach number flows. 
Also shown are the same quantities for the Lorenz model, which is discussed in
Section~\ref{lor}.

At any instant in time, the total convective kinetic energy is 
${1 \over 2}M_{CZ} v_{rms}^2$, where $M_{CZ}$ is the mass of the
convective zone and $v_{rms}$ is the rms velocity\footnote{More precisely,
$v$ is the solenoidal velocity in the convective region, with overall translational
velocity removed, so that $v_{rms}$ is its root-mean-square value.},
while the kinetic energy in the isotropic part of the turbulent flow field is
$E_{turb} = {1 \over 2}M_{CZ} u_{t}^2$, where $u_t$ is the turbulent velocity,
which we define as the isotropic part of the turbulent flow\footnote{This follows
\cite{amy09}, \S2.4, in which the transverse velocities are used to estimate
the isotropic component of the vertical velocity. Thus, we use
$u_t^2 = {3\over 2}(u_x^2 + u_y^2)= {3 \over 4}v_{rms}^2$, 
where $u_x$ and $u_y$ are the
velocity components perpendicular to the direction of the gravity vector,
i.e., the tangential velocities.
 The vertical velocity also contains a significant contribution 
 (${1 \over 4}v_{rms}^2$) from the non-isotropic flow of the largest eddies.
A more careful discussion of the velocities in terms of principal
orthogonal decomposition is in preparation.}.
The reader is warned that
division of the flow into ``turbulent'' and ``large scale" flow is useful 
but an oversimplification, so that the exact relative values of these two 
kinetic energies depend upon the algorithm used in their definition, 
but is of order unity \citep{amy09,ma11}. Consequently the precise
distinction does not change the qualitative picture.

The buoyancy power is the rate at which kinetic energy per unit mass
is increased by buoyant acceleration. 
If it is integrated over the space containing the
convection zone, we have $ q_{int} =\int_{CZ} q dr $, 
where $q = -g u' \rho'/\rho_0$ is the buoyant acceleration
times the turbulent velocity; $q_{int}$ has units of velocity cubed.

In Figure~\ref{fig1} the simulations show a phase lag of about 20 seconds between 
the peaks in buoyancy term ($q_{int}^{2/3}$) and turbulent kinetic energy. 
This is about half the time it takes the flow to transit 
a distance $\ell_{CZ}$, the depth of the convection zone. 
It also corresponds to an e-folding time for
turbulent kinetic energy decay due to Kolmogorov damping, where
$\varepsilon_K = u_{rms}^3/\ell_d = (u_{rms}^2/2)(2u_{rms}/\ell_d)$.
Power spectra for both variables peak at 89 seconds; an average 
transit time is $51\rm\ s$. 

Figure~\ref{fig1} shows multimodal behavior in the 3D simulations; 
preliminary results from a
quantitative analysis \citep{ma11}  using principal orthogonal decomposition (POD) indicates that a single dominant mode
has about 43\% of the kinetic energy, the first five modes have 75\%, and
90\% is reached with the eleventh mode.  There is a strong dominant mode but
also significant energy in several other modes; the modes  interact in a nonlinear
and dynamic way.

The buoyancy power is a large scale feature and is strongly anisotropic 
(plumes move vertically). 
The dissipation implied by the turbulence occurs at the Kolmogorov
scale (which is tiny); this dissipation is wide-spread in space \citep{amy09},
including the entire turbulent region on average, and bounded by the stably stratified layers. 
Because buoyancy and dissipation occur at vastly different
length scales, they are weakly coupled.

In low Mach number flow, on average over time, 
the buoyant driving must balance the turbulent damping for
a quasi-steady state  to exist
(see \cite{amy09} for a detailed discussion, especially their Eq.~33). 
The time averaged viscous dissipation is
\begin{equation}
\epsilon_{turb} = \overline{ q_{int} }/\ell_{CZ} \approx \overline{ v_{rms}^3 }/\ell_{d} 
\approx 1.54\overline{ u_t^3 }/ \ell_d, \label{eturb}
\end{equation}
where $\ell_{d}$ is the dissipation length, $\ell_{CZ}$ is the depth of the
convection zone, and the over-lines indicate an average over time (two turnover
times or four transit times,  about 200 seconds, in the analysis).

From Equation~\ref{eturb} it is clear that the turbulent dissipation is of third order
in the turbulent velocity, while the buoyancy power, $q = -g u' \rho'/\rho_0$,
is second order (because $\rho'$ is first order). This means that in the turbulent
regime there is a unique turbulent velocity which satisfies the condition that
buoyancy power balance turbulent damping. This has the nature of an eigenvalue
problem.
The assumption of  a turbulent cascade implies finite turbulent velocities, with  macroscopic structure. 
Equation~\ref{eturb} applies as an integral over a region much larger than the Kolmogorov scale; at the Kolmogorov scale
microscopic dissipation occurs by the Navier-Stokes term (\cite{ll59}, \S15). If we
dot this Navier-Stokes term with the velocity to generate a rate of kinetic energy destruction by viscosity (see Equation~\ref{ke-eq}), we have
\begin{equation}
({\bf u \cdot} \partial {\bf u} /\partial t )_{visc} =  \nu {\bf u \cdot} \nabla^2 {\bf u} \label{enavier}
\end{equation}
which does scale as the velocity to second order. When integrated over the
turbulent cascade, Equation~\ref{enavier} becomes Equation~\ref{eturb}. 
The extra factor of velocity comes from the rate at which turbulent kinetic
energy is delivered to the Kolmogorov scale.
Turbulence is an inherently
nonlinear process, which sets up its own scales and structures. Care must be
taken in analyzing it with techniques involving expansion in series 
\footnote{See \cite{am11b} for an example of the danger of using linear stability
analysis for turbulence in stars.}.

The dissipation is seen to be driven by the isotropic part of 
the flow $u_t$, which accounts for about
three-fourths of the kinetic energy according to estimates by  \cite{amy09} 
(which are roughly similar to those from POD analysis by \cite{ma11}). 
Note that $ v_{rms}^2 = {4 \over 3} u_t^2$ which gives
the vertical offset between the solid and dotted curves in Figure~\ref{fig1}.
Consider the average conditions from 200 to 800 seconds in the 
simulations.  The average level of buoyancy (actually $q_{int}^{2/3}$
is plotted) is higher than that of kinetic energy in isotropic turbulence. 
If we equate (1) the power generated by buoyancy, to (2) the dissipation due
to turbulence, both averaged over a few transit times, we find
\begin{equation}
    \ell_{d}/\ell_{CZ} = \overline{ u_{rms}^{3} } /\overline{ q_{int} }  \approx 0.85.
\end{equation}
Thus, in the \cite{ma07b} simulations, {\em the dissipation length is found to be
essentially the depth of the convective zone}, consistent with Kolmogorov 
theory.

These results are far more general than the particular stellar situation we have
discussed. While the neutrino cooling may seem exotic to some stellar evolutionists, in fact it behaves somewhat like the more familiar cooling by
radiative diffusion, and has no strange effect on the turbulence. We note that
the original simulations \citep{ma07b}, which included core hydrogen burning
cooled by radiative diffusion as well as oxygen shell burning cooled by neutrinos, explicitly showed this similarity.

In the long term,  the thermal state of the convection zone is supposed to  evolve
toward a global thermal balance between total heating by nuclear
burning and total cooling by neutrino emission \citep{wda72b,wda96}. 
This is illustrated in
Figure~\ref{fig2}, which shows time average luminosities of heating and of cooling,
as a function of the entropy of the convection zone. At lower entropy, as in the
3D simulation,
the heating is dominant, causing the entropy to increase. This is
accomplished primarily by expansion, with a small increase in temperature.
The decrease in density causes a bigger change in nuclear burning than in
neutrino cooling, so that a thermal balance would be attained (shown
as a vertical line) when the heating and cooling curves cross.

The gradual rise in turbulent kinetic energy in the simulation, shown in Figure~\ref{fig1},
occurs in a shell which is below the entropy of balanced heating and
cooling, so that heating dominates. The pulses are much faster than
this secular evolution, which changes on a time scale of $t \ge 10^3$~seconds.

\section{The Lorenz Solution\label{lor}}

The Lorenz model is a convective
roll, or cell, whose dynamics are described by three amplitude equations.
This is a simple example of an elegant method of reduction of turbulent flow
to a low-order set of dynamical equations using amplitudes of a proper orthogonal decomposition (POD) of numerical simulations or extensive experimental data sets
\citep{hlb96}. 

\begin{figure}
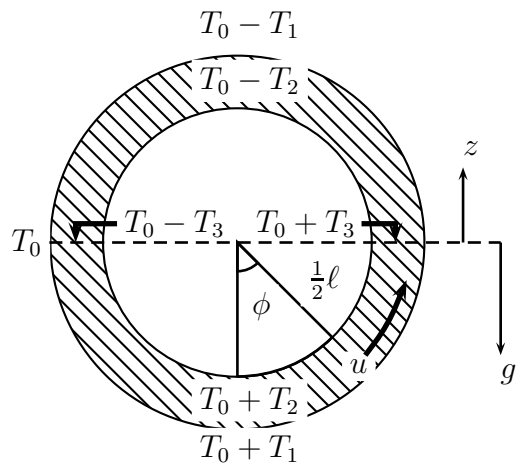

\figurenum{3}
\
\psset{unit=.5cm}
\pspicture*[](-12,-2)(4,12)
%\psgrid(0,0)(-12,-2)(4,12)
\pscircle[fillstyle=vlines](-5,5){5}
\pscircle[fillstyle=solid](-5,5){3.6}
\psline[linewidth=1pt]{<-}(2,2)(2,5)
\rput*[l]{N}(2,1.5){\large \bf $g$}
\psline[linewidth=1pt]{->}(1,5)(1,7)
\rput*[l]{N}(1,7.5){\large \bf $z$}
\rput*[l]{N}(-11,5){\large \bf $T_0$}
\rput*[l]{N}(-8,5.5){\large \bf $T_0-T_3$}
\psline[linewidth=2pt]{->}(-8.2,5.5)(-9.3,5.5)(-9.3,5.0)
\rput*[l]{N}(-4.5,5.5){\large \bf $T_0+T_3$}
\psline[linewidth=2pt]{->}(-1.7,5.5)(-0.8,5.5)(-0.8,5.0)
\pswedge[linewidth=1pt,fillcolor=lightgray](-5,5){3.6}{270}{315}
\rput*[l]{N}(-4.6,3.3){\large \bf $\phi$}
\pswedge[linewidth=1pt,fillcolor=lightgray](-5,5){0.8}{270}{315}
\rput*[l]{N}(-3.1,4.0){\large \bf ${1 \over 2}\ell$}
\rput*[l]{N}(-6,10.8){\large \bf $T_0-T_1$}
\rput*[l]{N}(-6,9.3){\large \bf $T_0-T_2$}
\rput*[l]{N}(-6,-0.5){\large \bf $T_0+T_1$}
\rput*[l]{N}(-6,0.8){\large \bf $T_0+T_2$}
\rput*[l]{N}(-2,1.7){\large \bf $u$}
%                                               q
\psline[linewidth=2pt,linearc=4.5]{->}(-1.6,2)(-1,2.8)(-0.5,4)
\psline[linewidth=1pt,linestyle=dashed]{-}(-10,5)(2,5)
\endpspicture
\caption{The Lorenz Model of Convection: Convection in a Loop.}
\label{fig3}
\end{figure}

The Lorenz model is a better mathematical representation of the dynamics of a convective cell than MLT in that the acceleration and deceleration over a convective cycle are integrated to determine the motion, rather than prescribed. Because it uses a single mode, the Lorenz model does not have sensitivity to local variation; it is a {\em global} model, unlike MLT, which is local. 
 Lorenz devised the model as a point of principle test of meteorological convection, which like the stellar
problem, is damped by a turbulent cascade. To make use of this extensive
literature \citep{cvit,frisch,gleick,thomp},
 we use the original version of \cite{lorenz}, with the same Prandtl and
Rayleigh numbers, to explore the implications on dynamics of the strange attractor.
The original Lorenz formulation is a 2D, low Mach number, and 
single mode model (equivalent to Figure~\ref{fig3}). 
A transition to 3D may be made using the solutions
of \cite{ch61}, but is not necessary for present purposes. 
Real convection \citep{libchaber} is expected to be 
single mode only near the onset of convective instability. 

\placefigure{3}

The Lorenz equations (see Appendix~A) are:
\begin{eqnarray}
dX/d\tau =& -\sigma X + \sigma Y    \label{lor1}\\
dY/d\tau =& -XZ +rX -Y \label{lor2}\\
dZ/d\tau =& XY - bZ, \label{lor3}
\end{eqnarray}
In the classical Lorenz model, $X$ is proportional to the speed
of convective motion, $Y$ is proportional to the temperature 
difference between ascending and descending flow, and $Z$ is 
proportional to the distortion of the vertical temperature profile from
adiabatic \citep{lorenz}. 
Here $\tau$ is a time in thermal diffusion units, $\sigma$ is the 
effective Prandtl number, $r$ 
is the ratio of the Rayleigh number to its critical value for onset of
convection, and 
$a$ is the ratio of the depth to the width of the convective cell,
so that $b=4 /(1 + a^{2})$; see \cite{lorenz}. 
The Prandtl number is the ratio of coefficients of the viscous dissipation term 
to the thermal mixing term. 
Lorenz\footnote{A justification is simply that this allows the three Lorenz equations to capture chaotic behavior. See Appendix B for a discussion of the
physical implications of the Prandtl number for stars.} took $\sigma = 10$, and chose the most unstable mode so that  $a^2 = 1 / 2$, 
and $b =8/3$. For this mode, and a Rayleigh number of
$r=470/19=24.74$ times the critical value,  steady flow becomes unstable.

Figure~\ref{fig3} illustrates the structure and notation in the Lorenz model.
The Prandtl number may be expressed as $\sigma = \tau_{rad}/\tau_{visc}$, the ratio of the radiative
cooling time scale to the viscous time scale. The viscous damping time is taken to
be constant, $\tau_{visc}=1/\Gamma$. 
The Rayleigh number, in units of its value at the onset of laminar convection, is
\begin{equation}
r = g \beta_T T_1 / \ell \Gamma K T_0,
\end{equation} 
where $K = 1/\tau_{rad} $,
$\beta_T = -(\partial \ln \rho / \partial \ln T)_P$, and $g$ is the gravitational acceleration.
The lapse rate of the adiabatic background $T_E$ is $g/C_P$, so $T_1 = g (\ell/2)/C_P$.
In astrophysical notation, 
\begin{equation}
 \nabla - \nabla_a = \nabla_a(T_2/T_1 -1 ),
 \end{equation} 
 where $T_2$ is the temperature variation in the vertical direction (Fig.~\ref{fig3}).
 The depth of the roll, in pressure scale heights, is
\begin{equation}
\ell/H_P = {1 \over \nabla_a} \ln [(T_0 + T_1)/(T_0-T_1)].
\end{equation}

\subsection{Energy fluxes\label{Afluxes}}
In the meteorological case
the fluid motion (wind)  is of primary interest, however the energy fluxes\footnote{In discussion of the Lorenz model, both heat flux and enthalpy flux are mentioned; which is correct? Depending upon the physics perspective, both may be. In the simplest Lorenz model, the fluid in strictly incompressible, so that the work term
$PdV$ is strictly zero, and in this case only heat content can carry energy, so heat flux is relevant. However, we may interpret the equations in terms of a stratified system with low-Mach number flow, having a stratification in density as well as temperature. Then $dV$ is not zero, $PdV$ work is done, and the relevant flux is the enthalpy flux.} provide a means to connect with stellar observations.

The vertical enthalpy flux due to radiative diffusion of the background is 
\begin{eqnarray}
F_E &=& -\rho \nu_T C_P dT_E/dz \\
&=& \rho \nu_T C_P T_1(2/\ell) ,
\label{frade}
\end{eqnarray}
which is constant in the Lorenz model.  The thermal conduction coefficient is
$\nu_T = (4acT^3/\rho \kappa)/(\rho C_P)$ 
and $K = \nu_T(1/\ell)^2$.
Here $ z = - {\ell\over 2} \cos \phi$, the radius of the roll being $\ell /2$.
Because the background temperature is taken to be linear in $z$, the divergence
of this flux is identically zero, and therefore gives no local heating or cooling.
The additional vertical enthalpy flux, due to the temperature perturbation, is 
\begin{eqnarray}
F_z &=& -\rho \nu_T C_P d(T-T_E)/dz \\
 &=& \rho \nu_T C_P (T_2 -T_1)(2/\ell) .
\label{fradza}
\end{eqnarray}
If we define a potential temperature\footnote{Here potential temperature is used
as in fluid dynamics and meteorology \citep{dutton,tritton}; the potential
temperature is measured relative
to an adiabatic background, and may be small even if the convective region is
strongly stratified (as is often the case in stars).}
$T_4 = T_1 - T_2$, the vertical flux is separated
into two parts, so
that the adiabatic background value is denoted by $T_1$ and the changes
caused by motion are contained in $T_4$.

The flux in the horizontal direction is 
\begin{eqnarray}
F_y &=& - \rho \nu_T C_P dT/dy \\
&=& - \rho \nu_T C_P T_3(2/ \ell),
\label{frady}
\end{eqnarray}
which averages to zero by symmetry; here $y = {\ell \over 2} \sin \phi$.
Both $F_z$ and $F_y$ are proportional to a potential temperature which varies
in time. 

The net vertical enthalpy flux  will not be constant in height $z$, so that its divergence is nonzero. This 
implies local heating/cooling, which would have to be compensated for in a steady
state, and is a consequence of considering only a single mode. 
Smaller scale modes would be needed to
deal with the local heating/cooling; see \cite{cm91,can96} for efforts to include
a full spectrum of modes.  
To get the average flux through the cell for a single mode, we take the double projection, of $u$ and of $T$ on the vertical direction,
\begin{eqnarray}
F_e(\phi) &=& \rho  u_z C_P T \nonumber\\
&\rightarrow& \rho C_P u T_3 \sin^2 \phi. \label{Fcphi}
\end{eqnarray}
The enthalpy flux is zero at the top and bottom of the roll
($\phi = \pi$ and $0$),
and a maximum at the midpoint ($\phi = \pi/2$). The average value of $\sin^2\phi$
over the roll, $\phi = 0$ to $2 \pi$, is $0.5$. This is to be compared with the $\alpha_E$ of
\cite{amy09}, which was the range $0.68$ to $0.85$ for the simulations then available.
Because the simulations are multi-mode, this slightly higher value for the correlation seems natural.

Integrating over the cell, $0 \leq \phi \leq 2 \pi$, gives
\begin{equation}
F_e = {1 \over 2} \rho C_P u T_3, \label{fenthalpya}
\end{equation}
which is the vertically averaged enthalpy flux of the cell.
This is the ``convective flux" in stellar models. The up-flow enthalpy flux equals the
down-flow value, and both are positive (hot up-flows and cold down-flows both transport energy upward).

The ratio of enthalpy flux to the total radiative flux ($F_z + F_E$) is
\begin{equation}
F_e/F_r = ( \ell/ 4 \nu_T )( u T_3 /T_2 ),
\end{equation}
which is a function of time.

The kinetic flux exactly cancels in this formulation, as in MLT, with the up-flow being 
the negative of the down-flow. This is {\em not} true in the simulations.
The kinetic flux does approach zero for convective regions which are
so thin that they are almost unstratified. However, 
stratified convection zones have an asymmetry in up and down flows,
giving a modest net kinetic energy flux (downward for pure top driving,
up for pure bottom driving, and both for more general cases, \cite{ma09}).

We may express fluxes in units of the radiative flux of the background,
$F_E $, and in terms of the variables in the Lorenz equations.
The excess vertical radiative flux is
\begin{eqnarray}
F_z/F_E &=& T_4/T_1 \nonumber\\
&=&  Z/r,
\label{fradz}
\end{eqnarray}
where  $T_1 = g \ell/2 C_P$,  $T_0 = g H_P/ (C_P-C_V)$ and
$\gamma = C_P/C_V$, and  
\begin{equation}
r = {\gamma - 1 \over 2\gamma}\Big ( {g Q \over H_P K \Gamma}\Big ).
\end{equation}
The horizontal radiative flux is
\begin{eqnarray}
F_y/F_E &=& T_3/T_1 \nonumber\\
&=&  Y/r.
\end{eqnarray}
The enthalpy flux of the cell, in the vertical direction, is
\begin{eqnarray}
F_e /F_E =  XY/2r.
\label{fenthalpy}
\end{eqnarray}
In general, without need for any additional mechanism to cause variability, 
{\em the net flux of energy through a turbulent cell varies with time.}

\subsection{Steady-state solutions}
A steady state solution for the Lorenz equations is
\begin{equation}
X = Y = Z = 0,
\end{equation}
and, for $r \ge 1 $, a second solution, for a stable convective roll, appears:
\begin{equation}
X=Y= \pm[b(r-1)]^{1/2}; Z = r-1.
\end{equation}
The second solution is unstable if $\sigma >2$ and $r > r_c$ where
\begin{equation}
r_c = \sigma(\sigma +b +3)/(\sigma -b -1),
\end{equation}
and this instability gives rise to fluctuations in velocity \citep{lorenz}. The instability also gives fluctuations in energy flow through the cell.

The steady state solutions might be of practical value for stellar evolutionary codes,
to the extent that they provide an estimate of the average behavior of the convective
cell. However, the fluctuations are large (in no sense are they "small perturbations"),
and may cause nonlinear complications. For example, a thermonuclear runaway would
be sensitive to the largest value of the temperature fluctuations because that would
affect the burning rate \citep{am11b}.  The fluctuations drive entrainment episodes, and affect the
mixing of composition \citep{ma07b}.
The fluctuations may be able to modify the driving of 
pulsations in stars with vigorous convection zones. In red supergiants, coupling of
turbulent fluctuations in the surface convection zones to both radial and non-radial
pulsations seems likely (see below). 

\subsection{Non-steady Solutions}
In order to compare the Lorenz model to 3D simulations, it is desirable to have
comparable starting conditions. Arbitrary choices can give large initial transients before the attractor controls the behavior. The 3D simulations start with zero
velocity but finite temperature deviations from an adiabatic gradient. We take
\begin{equation}
X=0, Y=[b(r-1)]^{1\over 2}, Z= r-1,
\end{equation}
as initial conditions,
which sets the velocity to zero and uses steady state values for the temperature fluctuations.

Figure~\ref{fig1} shows the behavior of the Lorenz model of convection
(panel labeled LORENZ),
for a similar number of pulses as shown for the 3D simulation, 
and the same variables: buoyancy power $XY/(b(r-1))$ to the two-thirds
power, and kinetic energy per unit mass $X^2/(b(r-1))$. 
The factors $b(r-1)$ are chosen so that the steady state values are
normalized to unity, but are qualitatively correct for comparison 
with the 3D simulations. 
The velocities in Fig.~\ref{fig1}  are not precisely
the same, the 3D simulations giving turbulent velocity while the Lorenz
speeds are more appropriate to  ``coherent structures". They are related (see \S~3.1), as the large scale velocities become unstable and turbulent, and their kinetic energies are the same order of magnitude \citep{amy09,ma11}.

The time is measured in the dimensionless units of Lorenz; a turnover
time is two transit times, and roughly the time between peaks.
We see that the peaks in buoyancy power slightly precede those in 
kinetic energy, as in the 3D simulations but less dramatically. This difference
is related to the fact that the Lorenz model has viscous dissipation acting directly
on the large scale velocities, while the 3D simulations have dissipation at the
Kolmogorov scale, which is separated from the large scale by the turbulent
cascade, involving many, many reductions in length scales 
(see Section~\ref{cascade}).
Additional modes would fill in the ``valleys" in the Lorenz model (see \S\ref{Scell}).
Over the time shown, the average kinetic energy is 0.968 of
the formal steady state solution, which is
shown in Figure~\ref{fig1} as the dashed horizontal line in the Lorenz panel. 

The degrees of freedom in the three-dimensional simulations, and in the Lorenz 
model, are dramatically different.  The
floating-point operation count differs by a large factor:  $\sim10^8$ (several times
$10^8$ zones times 7 effective scalar variables for simulations, versus
three amplitudes for the Lorenz model). With such an extreme compaction,
it is striking that they give a similar picture for fluctuations in convection.

\subsection{Long Term Behavior}

\begin{figure}
\figurenum{4}
\includegraphics[angle=-90,scale=0.3]{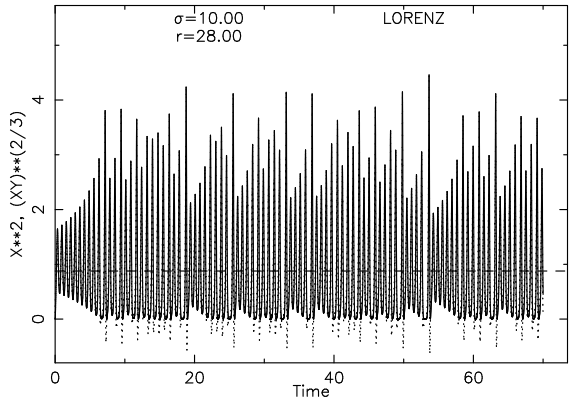}
\caption{Later Behavior of the Lorenz Model of Convection,
for $\sigma=10$, $r=28$, and $b=8/3$. 
The format is the same as Figure~\ref{fig1}, but the time interval
is longer by a factor of 10. Chaotic behavior is beginning to
appear, with fluctuations in kinetic energy and buoyant driving
sometimes exceeding the steady state value by a factor of 4.
}
\label{fig8}
\end{figure}
\placefigure{4}

Unlike the full 3D simulations, it is trivial to extend the Lorenz model to
later times. Figure~\ref{fig8} shows an extension in time by a factor of 10.
There is a relatively steady growth in amplitude up to time $\sim$7,
at which point chaotic behavior begins to appear. The fluctuations in
kinetic energy (and 
$q_{int}^{{2 \over 3}}$) are large, sometimes exceeding the
steady state value ($\sim$1) by a factor of 4. 
The convective luminosity from a single Lorenz model shows the same qualitative
behavior.

This drastic behavior raises two interesting possibilities:
\begin{enumerate}
\item  The numerical simulations will follow the solutions of the
Lorenz equation, and exhibit vigorous and chaotic fluctuations at
later times, or,
\item The multimode behavior of the simulations will allow even stronger dissipation,
preventing extreme behavior.
\end{enumerate}
Either way, the result is important for the evolution of
supernova progenitors, especially regarding the effects of fluctuations
on mixing (yields) and outbursts.
Simulations of multiple burning shells (C, Ne, O and Si) in 2D  have been continued further than the 3D simulations for the oxygen burning shell. They appear to
follow the first option, so that the prediction from 2D simulations is that core collapse progenitors will have violent eruptions prior to core collapse 
\citep{am11b}. Full 3D simulations need to be extended to later times at which
the Lorenz model predicts chaotic behavior.

\subsubsection{Duration}
The oxygen shell might not last for the $\sim$70 transit times shown in Figure~\ref{fig8}; 
a linear estimate suggests that it consumes its fuel in about 100 traversal times 
(the Damk\"ohler number is $ D_a \le 0.01$, \cite{amy09}), so the background 
evolution should not be neglected for such time intervals. Oxygen burning is likely to
be a more dynamic event than previously supposed (e.g.,  by \cite{whw02}).
Averaging over multiple cells may moderate the net fluctuation (see below).

For convection zones in other stars, the number of traversal times available may
be much larger. The sun has a deep surface convection zone (20 pressure scale
heights); a plume would, if unimpeded, fall through the
convection zone in about 2 hours;  it would take about 7 centuries to
process all the mass of the convection zone through the surface layers. These
times bracket the effective mixing time, so that of order $10^7$ to $10^{13}$ mixing
times would occur over the age of the present-day sun.

\subsubsection{Chaotic Behavior}

\begin{figure}
\figurenum{5}
\includegraphics[angle=-90,scale=0.3]{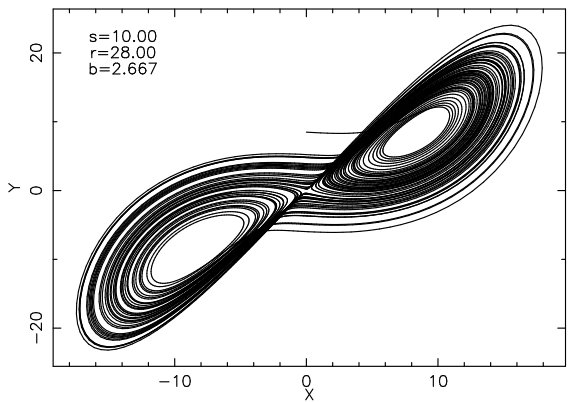}
\caption{Long Term Behavior of the classic Lorenz Model for X (speed) and
Y (horizontal temperature fluctuation),
for $\sigma=10$, $r=28$, and $b=8/3$. 
The trajectory is shown for 70 Lorenz time units. Both X and Y switch signs, 
but there is an overall correlation (hot zones rise while cool zones sink). 
This correlation gives rise to positive values for buoyancy flux and  enthalpy flux.
}
\label{figxy}
\end{figure}
\placefigure{5}

\begin{figure}
\figurenum{6}
\includegraphics[angle=-90,scale=0.3]{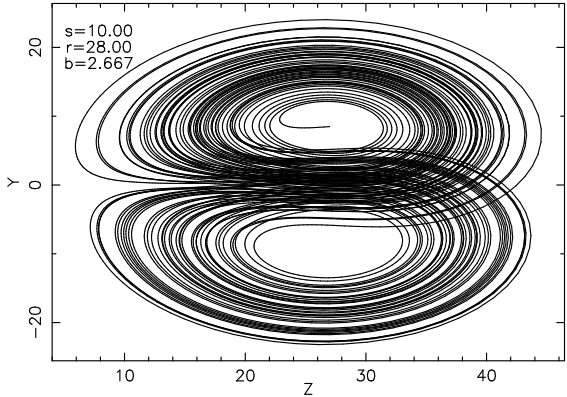}
\caption{Long Term Behavior of the classic Lorenz Model,
for $\sigma=10$, $r=28$, and $b=8/3$. 
Same as Figure~\ref{figxy}, but with Y and Z (vertical temperature
fluctuation) shown. Z is a positive for these parameters. 
Y has both positive and negative steady state values, about which
it orbits. 
}
\label{figyz}
\end{figure}
\placefigure{6}

Figures~\ref{figxy} and \ref{figyz} show the familiar long term behavior of the Lorenz model
\citep{gleick}.
Figure~\ref{figxy} shows the trajectory in X and Y space for the first 70 dimensionless
time intervals.  X is the dimensionless speed and Y the dimensionless temperature
variation in the horizontal direction.
Both X and Y switch signs, but there is an overall correlation because
there is a net enthalpy flux, which is proportional to XY. 
The correlation is in the sense
that hot regions rise and cool regions sink.  The radiative flux associated with the
horizontal temperature variation is proportional to Y, and while on average over
long times is zero, it achieves this in segments of  time in which first one then the other
direction (sign) are favored. This is the strange attractor at work.

Figure~\ref{figyz} shows the same time
interval, but plots Y  and Z (the dimensionless temperature in the vertical direction,
parallel to the gravity vector). Notice that Z has positive values about which it orbits.
The stratification breaks the symmetry in vertical versus horizontal directions.
The radiative flux in the vertical direction has a steady part and a fluctuating part.
The latter is proportional to Z, and is positive (outward energy flow). 

In MLT, the Schwarzschild discriminant (${\cal S} \equiv \nabla - \nabla_{a}$) is
the temperature excess above adiabatic; this excess is assumed to be proportional to the enthalpy flux.
This is inconsistent with the Lorenz model, because the velocity X can have both
signs while Z does not. The error arises because in MLT the speed of convection
is taken to be intrinsically positive (to avoid this problem), and may be traced back
to a lack of conservation of mass ("blobs dissolve" into the environment rather than
flow back to complete a cycle).

\begin{figure}
\figurenum{7}

\includegraphics[angle=-90,scale=0.3]{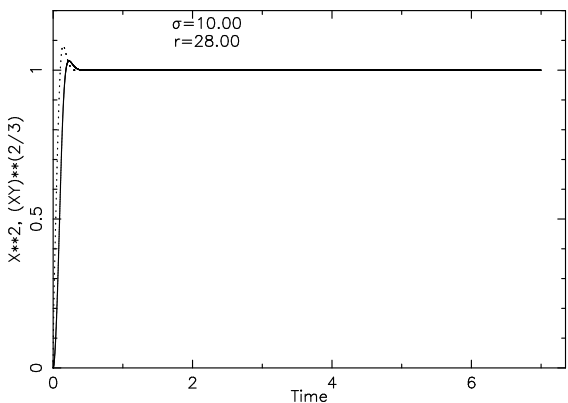}

\caption{Initial Behavior of the two Equation Model of Convection,
for $\sigma=10$, $r=28$, and $b=8/3$. The critical value for
instability of the convective roll is $r_c = 24.74$.
The format is the same as Figure~\ref{fig1}. The turbulent fluctuations are
eliminated, as they are in MLT.
}
\label{fig11}
\end{figure}
\placefigure{7}

\subsubsection{MLT}
What happens if we reduce the three equations of Lorenz
to two, forcing one variable to be at its steady state value? This is similar to
the MLT approach (assuming the mixing length parameter is chosen so that
the kinetic energy scale is physically correct; see \cite{amy10}).
Enforcing the steady state value for the vertical temperature excess, 
$Z = r-1$, but allowing $X$ and $Y$ to vary,
corresponds to a model with a single temperature variable (like MLT).
Such integrations are shown in
Fig.~\ref{fig11}. There are no pulses in kinetic energy or buoyancy; the curves
quickly approach a constant value. Convection proceeds by steady
motion in a roll;  a finite XY is required to give torque to make
the roll. This "two equation model" no longer has
a strange attractor; the pulses have been eliminated. This explains why stellar
evolutionary calculations which use MLT do not show these fluctuations\footnote{
In analogy, the two-body  (Kepler) problem in celestial mechanics is well behaved, while the unrestricted three-body problem is far more complex \citep{poincare}.}.

\section{Cells and Shells\label{Scell}}

\begin{figure}
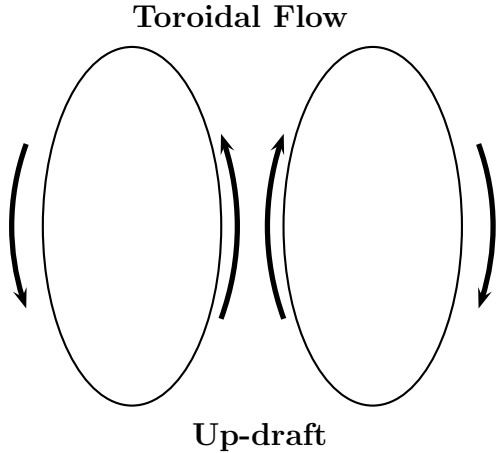

\figurenum{8}
\
\psset{unit=.4cm}
\pspicture*[](-9,-8)(9,8)

\psarc[linewidth=2pt]{->}(0,0){8}{160}{200}
\psellipse[fillstyle=solid](-4,0)(3,6)
\psarc[linewidth=2pt]{->}(-9.5,0){9}{-20}{20}

\psarc[linewidth=2pt]{<-}(9.5,0){9}{160}{200}
\psellipse[fillstyle=solid](4,0)(3,6)
\psarc[linewidth=2pt]{<-}(0,0){8}{-20}{20}

\rput*[l]{0}(-2,-7){\large \bf Up-draft}
\rput*[l]{0}(-4,7){\large \bf Toroidal Flow}

\endpspicture
\caption{The Lorenz Model extended: convection in a sphere composed of 
a cell of toroidal shape. With no rotation to break the symmetry, the
direction of the upflow vector is not constrained, and will be chaotically chosen.}
\label{fig5}
\end{figure}
\placefigure{8}

\begin{figure}
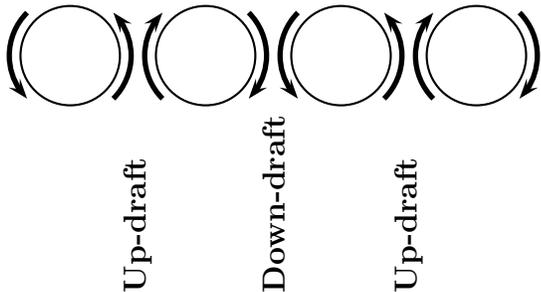

\figurenum{9}
\
\psset{unit=.45cm}
\pspicture*[](-12,0)(4,10)
\psarc[linewidth=2pt]{->}(-10,8){1.8}{135}{225}
\pscircle[fillstyle=solid](-10,8){1.5}
\psarc[linewidth=2pt]{->}(-10,8){1.8}{-45}{45}

\psarc[linewidth=2pt]{<-}(-6,8){1.8}{135}{225}
\pscircle[fillstyle=solid](-6,8){1.5}
\psarc[linewidth=2pt]{<-}(-6,8){1.8}{-45}{45}

\psarc[linewidth=2pt]{->}(-2,8){1.8}{135}{225}
\pscircle[fillstyle=solid](-2,8){1.5}
\psarc[linewidth=2pt]{->}(-2,8){1.8}{-45}{45}

\psarc[linewidth=2pt]{<-}(2,8){1.8}{135}{225}
\pscircle[fillstyle=solid](2,8){1.5}
\psarc[linewidth=2pt]{<-}(2,8){1.8}{-45}{45}

\rput*[l]{L}(0,1){\large \bf Up-draft}
\rput*[l]{L}(-4,1){\large \bf Down-draft}
\rput*[l]{L}(-8,1){\large \bf Up-draft}

\endpspicture
\caption{The Lorenz Model extended: Convection in a shell composed of cells. Notice the alternation of the sign of rotation. This may be thought of as a cross sectional view of infinitely long cylindrical rolls, or of a set of toroidal cells, with pairwise alternating vorticity (see \cite{ch61}, \S16). Each cell may exhibit independent fluctuations in time and space.
}
\label{fig4}
\end{figure}
\placefigure{9}

\subsection{Multiple Modes in Cascade\label{cascade2}}

%\begin{deluxetable}{lrrrr}
%\tablecaption{Simple Richardson Cascade\label{table1}}
%\tabletypesize{\small}
%\tablewidth{210pt}
%\tablehead{ \colhead{$f$} &
%\colhead{$f^{2/3}$} & \colhead{$n=1$}  & \colhead{$n=2$} & \colhead{$n=3$} 
%}
%\startdata
%$1/e$ & 0.513 & 0.487 &  0.250 & 0.128  \\
%$1/2$ & 0.630 & 0.370 & 0.233 & 0.147 \\
%$1/\sqrt{2}$ &  0.707 & 0.293  & 0.207  & 0.146 \\
%\\
%POD\tablenotemark{a} & \nodata  & 0.438 & 0.131 & 0.0875 \\
%\enddata
%\tablenotetext{a}{\cite{ma11}.}
%\tablecomments{See \S\ref{cascade}.}
%\
%\end{deluxetable}
%
%\placetable{1}

Figure~\ref{fig1} shows that the primary difference in the Lorenz model and
the 3D simulations is that the Lorenz model has only a single mode, while
the simulations are  obviously multimodal. This difference may be superficial. The Lorenz model in this application (as in the original  meteorological one) has additional modes implied by the turbulent cascade which mediates the damping (i.e., they are
implicitly in $\Gamma$; see Appendices \S\ref{viscous} and \S\ref{prandtl}).
A simple Richardson cascade was discussed in \S\ref{cascade}, in which  $f$, the fractional change in length scale for each step in the cascade, is assumed to be
a constant \citep{frisch,davidson}. This is not very plausible for the largest scale modes because they are the most sensitive to boundary conditions (they must fit into the convective region), but is simple and instructive.
The fractional time spent in the cascade for each mode may be shown to be the fractional kinetic energy in that mode. 
Using \S\ref{cascade}, this gives $ f^{(2/3)(n-1)}(1-f^{2/3})$ for $n=1, 2, \cdots$, or roughly 0.37, 0.23, and 0.14 for the first three.
There is a dominant mode accompanied by several weaker but significant ones. 

One way to proceed would be to introduce additional modes into the Lorenz model  (chosen with guidance by the 3D simulations), and to generate an expanded set of amplitude equations which generalize the three of Lorenz  (Eq.~\ref{lor1}, \ref{lor2}, and \ref{lor3}). This would make the system multimodal, and allow for modal interactions. This approach is a simplified version of an elegant proposal by Lumley and collaborators \citep{hlb96}: empirical eigenfunctions are constructed by POD from simulations, introduced into the differential equations  to derive a set of nonlinear ODE's for the amplitudes of each mode (a Galerkin projection), and this set of ODE's is solved to generate the evolution of the average properties of the turbulence. This is being explored \citep{ma11} as a way to  capture more fully  the dynamics and multimodal behavior seen in the 3D simulations.

\subsection{Multiple Cells}
Suppose we envisage the convective region to be populated by cells, each of which is a separate Lorenz model representing the largest mode, a convective roll, in that cell. 
A special case is the convective core:
 the geometry is not that of a layer, but a sphere\footnote{\cite{wpj03} found such a "giant dipole" behavior in their 3D simulations of almost fully convective spheres.}, as is indicated in Figure~\ref{fig5}.
 The flow has a toroidal structure.
 In this case the gravitational acceleration goes to zero at the origin (the center of mass of the star), but the velocity need not be zero there. 
 If there is no net rotation, then the direction of the up-flow is undetermined, and will be chosen chaotically by the turbulent flow. 

For a convective shell, we imagine that the layer is filled by cells, as
shown schematically in
Figure~\ref{fig4}. For the oxygen burning 
shell, the inner and outer radii are about $4 \times 10^8$ and $8 \times 10^8$~cm,
respectively. The area of the spherical shell, evaluated at the midpoint in
radius, is $4\pi (6\times 10^8)^2$, and the cell is taken to be roughly square,
so that its area is  $\ell^2 = (4 \times 10^8)^2$. 
The ratio of shell area to cell area is about $9\pi$, so we assume there are 
roughly $~30$ cells spread over the spherical shell. In general, the number of cells 
in a shell will depend upon the geometry of the convection zone.  

If the cells are not synchronous\footnote{See \cite{asy96} for a discussion of
synchronization of chaotic orbits.}, but act independently, the effect of the pulses  
will be smoothed when summed over the whole shell. However, the cells
may interact constructively; 
the solution to this more complex problem remains open\footnote{
Figure~23 in \cite{ma07b} suggests the complexity of the cell interaction within
a single convective region. 
The original simulations were on a wedge, of $27^o$ in theta and in phi.
Simulations with larger aspect ratio (larger angle wedges) do show a moderation 
of the total fluctuation, in qualitative agreement with the discussion above.
Figure~2 in \cite{ma06} indicates the complexity of interaction between multiple
burning regions even in 2D.}. At issue are both the interactions between cells in a
single convective region, and interactions between multiple convective regions
associated with different burning shells.
 
 The individual cells exhibit fluctuations not only in time, but also in space. Each cell 
 represents a mode which is unstable, and destroys itself and reforms, usually somewhere else. Because the
 medium is fluid, the pattern of cells is much more dynamic and less regular than
 that of a crystalline solid, so the Fig.~\ref{fig4} should be interpreted as representing
 a snapshot of a system which fluctuates in both space and time. 
 Hinode/SOT observations of the solar surface
 \citep{berger10} reveal such a highly dynamic, complex structure\footnote{We thank 
 Dr. A. Title for providing a copy of his spectacular movies of data and simulations.}.
 
\subsection{Irregular Variables}

\begin{figure}
\figurenum{10}

\includegraphics[angle=-90,scale=0.3]{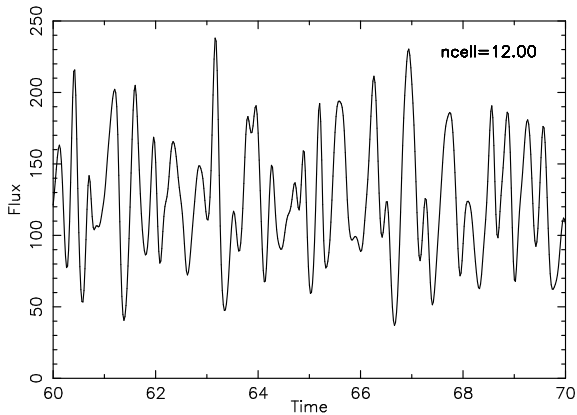}

\caption{Fluctuations of Luminosity in Convective Layer of 12 Cells of Random Phase,
for $\sigma=10$, $r=28$, and $b=8/3$. 
The dimensionless flux (luminosity) is shown for a convective layer with 12 visible
Lorenz cells. The luminosity variations are large and seemingly chaotic, suggestive
of irregular variables and Betelgeuse in particular \citep{kiss06}.
}
\label{fig12}
\end{figure}
\placefigure{10}

Figure~\ref{fig12} explores an idea of Martin Schwarzschild \citep{ms75}, who
estimated the number of convective cells in the sun and in red giants and supergiants.
He argued that only a modest number of cells
($\sim$12) would exist at any given time in a red giant or supergiant. 
To illustrate the point, we use the Lorenz model to approximate the behavior of a 
convective cell. We have estimated the convective flux for 12 cells at
random phase, by adding 12 time sequences from an computation
like Figure~\ref{fig8} (but extended to t=800), starting at 12
randomly chosen times in this interval. This time sampling is intended 
to approximate a spatial ensemble average.
Their flux, summed and normalized, is shown as a solid line, for a dimensionless time
from 60 to 70, which corresponds to roughly 20 pulses. The signal is noisy and
looks ``chaotic".

\begin{figure}
\figurenum{11}

\includegraphics[angle=0,scale=0.3]{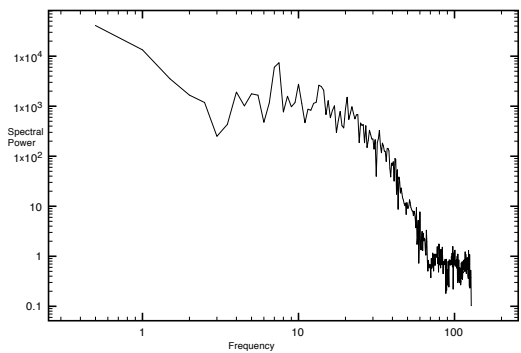}

\caption{Power spectrum of the luminosity fluctuations implied by turbulence alone.
There is no sharp peak, but a broad distribution of power, as would be expected from
a chaotic source. Resonant interaction with normal modes of the star could add peaks
to the spectrum,
which would provide an observational constraint on interior structure, analogous to
astro-seismology of solar-like stars.
}
\label{fig13}
\end{figure}
\placefigure{11}

These fluctuating cells make up a convective region, and will couple to the normal
modes of the star to cause both radial and non-radial pulsations. The amplitude of these pulsations will depend upon the overlap integrals between the normal modes and the cell motion, and the stellar damping. This suggests that the noisy behavior will be combined with the relatively cleaner periodicity of the normal modes, giving a power spectrum with a base like that shown in Figure~\ref{fig13}, but with superimposed spikes corresponding to the excited normal modes. 
While turbulent convection alone is sufficient to cause luminosity fluctuations, it occurs in regions of high opacity and partial ionization, which also drive pulsation, so that composite behavior and multiple periods may be expected.

Joel Stebbins (pioneer of photoelectric astronomy) 
monitored the brightness of Betelgeuse ($\alpha$ Orionis)
from 1917 to 1931, and concluded that "there is no law or order in the rapid 
changes of Betelgeuse" \citep{gold84}, which seems apt for Fig.~\ref{fig12} (the
Lorenz strange attractor) as well. More modern observations \citep{kiss06} show
a strong broad-band noise component in the photometric variability. 
The irregular fluctuations of the light curve are aperiodic, 
and resemble a series of outbursts.
Direct 3D simulations of Betelgeuse \citep{chia10} show the same complex
behavior (and have the advantage that they predict the detailed spectral behavior
as well). This should be no surprise; the 3D equations have embedded in them
a strange attractor. 

Increasing the
number of cells reduces the level of fluctuation about the mean.
Averaging over the 2 million granules of the sun gives a very stable luminosity, which
would plot as a straight line in Fig.~\ref{fig12} (even 2000 random cells do this).
However, the size of the cells at the bottom of the solar convection zone will be larger
(hence fewer cells), and if chaotic might give a long term modulation to the solar 
luminosity. Full star simulations of the whole solar convective zone, with sufficient
numerical resolution to give well developed turbulence, should shed light on this issue. 

Application to turbulent stellar atmospheres requires
surmounting two difficulties: (1) the flow is no longer low-Mach number (see \S~2.2),
and (2) the ionization zone causes dramatic changes in opacity (assumed constant
in the Lorenz model). Fortunately 3D atmospheres exist, and analysis such as we have
done on stellar interior convection simulations is feasible.

While this paper was in preparation, \cite{stothers} re-examined the idea
that giant convective cell turnover is the explanation of the long secondary period
observed in semiregular red variable stars \citep{sl71}, including Betelgeuse and
Antares. Stothers used MLT to derive a velocity scale for the overturn, also relying on general features\footnote{\cite{amy09} discuss and contrast these and other simulations.} of simulations of \cite{cs96}; see \S3 in \cite{stothers}. 
This theory appears to work directly as a complement to the discussion above.  
The use of a Lorenz model for the giant cells already implies real dynamics. 
The strange attractor necessarily provides  variability in luminosity, with its own quasi-period \citep{wok04} and  velocity scale (see \S4 in \cite{stothers}). 
Several giant cells\footnote{This idea can be tested observationally and numerically; the structure of the convection region will constrain the number of cells, which may be compared with the amplitude of  luminosity fluctuations.}  are at work, each with a quasi-period of order of the transit time,
and therefore similar to the estimate of Stothers, and the observations. 
The larger convective velocities needed are simply a necessary consequence of the dynamics implied by the convective luminosity \citep{amy10}. 

The introduction of turbulence as an {\em active} agent in the discussion of stellar variability (e.g., \cite{stothers,wok04}) seems timely.
An interesting improvement would be to use POD empirical
eigenvalues from simulations (e.g., \cite{chia10}, \S~\ref{cascade2} above), and develop a low order
dynamical model to explore long duration behavior.

\section{Summary}

We have identified a major new feature of stellar physics: 
chaotic behavior due to turbulent fluctuations in stellar convection,
and corresponding luminosity fluctuations. 
While the simulations
upon which the  analysis was based were fully compressible, the theory
uses the approximation of sub-sonic flow.
Both numerically and analytically, a strange
attractor like that of \cite{lorenz} seems to appear naturally in stellar convection.

As a first approximation to more rigorous analysis, we have applied the Lorenz model to kinetic energy fluctuations in the oxygen burning shell, to the turbulent
energy cascade, and to fluctuations in luminosity in irregular variables. 

Figure~\ref{fig1} shows a comparison of the turbulent kinetic energy fluctuations in 3D simulations of turbulent flow and in the Lorenz model. No parameters
were adjusted to give a fit.
Additional modes, appropriate for turbulent flow, would improve the
comparison further.

This suggests  a new, {\it inherently nonlinear} mechanism for variability in stars,
the $\tau$-mechanism, which is caused by luminosity fluctuations
{\em directly} associated with turbulent convective cells. 
Because the mechanism is nonlinear, it is not captured by linear stability
analysis, which is a mainstay of variable star theory.
Such luminosity fluctuations  may have been observed already in the broad-band noise seen in
Betelgeuse (alpha Orionis \citep{gold84,kiss06}, and in the long secondary
periods in pulsating AGB stars \citep{wok04,stothers},
and are expected 
to be observable in principle in all stars with extensive surface convection zones,
including those with ``solar-like" variability. 
This mechanism is probably the cause of the strongly driven pulsations found
by \cite{wpj03} in their ``red giant" model; the development of those large pulsations was a clue which may now be more fully understood.

Such fluctuations provide a source of perturbations for instabilities,
and may induce mixing not presently accounted for in stellar evolutionary calculations. 
The fluctuations in convective velocity are comparable to average values.
If these fluctuations couple to nuclear burning, as for example in cases of
degenerate ignition, shell flashes, or later stages of oxygen and silicon 
burning\footnote{For new developments since this paper was submitted, see \cite{am11b},
where it is suggested that core collapse progenitors are dynamically active prior to
collapse.},
outbursts may develop.

\begin{acknowledgements}
This work was supported in part by NSF Grant 0708871 and 
NASA Grant NNX08AH19G at the University of Arizona,
and by the CLEAR sub-contract from University of Michigan. 
We  wish
to thank Fr. J. Fu\~nes (Specalo Vaticano), 
Prof. R. Ruffini (ICRAnet), and Prof. J. Lattanzio
(Monash),
P. Wood (Australian National University), and the
Aspen Center for Physics for their hospitality, 
Prof F. Timmes and S. Starrfield for discussions,
R. Stothers for helpful email, and the second anonymous referee for constructive
comments.

\end{acknowledgements}

\eject

\appendix

\eject

\section{APPENDIX: Physical Basis of the Lorenz Model}

 \noindent{\bf Mass conservation.}
 Conservation of mass is enforced by use of a stream function (\cite{ll59}, \S9); the simplest solution is a two dimensional, cylindrical "roll" (\cite{ch61}, p.~44).
The flow is assumed to be subsonic.

\noindent{\bf Momentum conservation.\label{Amomcons}}
The Navier-Stokes equation  may be written as
\begin{equation}
    D{\bf u}/Dt =  \nu {\bf   \nabla^{2} u} -{\bf g } \Delta \rho / \rho,
    \label{eqb1}
\end{equation}
in the low Mach number limit;
the last term is the buoyant acceleration. The flow executes a circle of radius $\ell/2$; see Figure~\ref{fig3}. 
The density fluctuation is
related to the temperature fluctuation by $\Delta \rho /\rho_0 
= \beta_T  (T-T_0)/T_0$,
where $\beta_T = - (\partial \ln \rho /\partial \ln T )_P$.

We separate the variable $\bf u$  by considering the flow to be a mass
flux which is constant around the ring at any given moment, represented by an
average speed $u$ which is a function of time only\footnote{
Variations in density due to change in hydrostatic stratification could be included
as well as the implied changes in velocity and cross-sectional area; 
see \cite{tritton}, p.~188-196).}.
The hydrostatic background in the convective region has  an
entropy $S$ which is constant. Then, 
\begin{equation}
{1 \over \rho}{ dP \over dr} = {dW \over dr} = -g,
\end{equation}
where $W= E + PV$ is the enthalpy per unit mass, and 
$W= C_PT$, and $C_P = {\gamma \over \gamma -1}{\cal R}Y$. 
Now $\int gdr = g \Delta r$ if $g$ is constant.
Lorenz assumed a linear temperature decrease with
height, which corresponds to constant gravitational acceleration $g$.
Using a height $z = (\ell/2) \cos \phi$,
\begin{equation}
W(\phi) = W(0) + g (\ell/2) \cos \phi,
\end{equation}
or
\begin{equation}
T(\phi) = T_0 + {g (\ell/2) \over C_P} \cos \phi.
\end{equation}
We denote $T_1 = g (\ell/2)/C_P$, so
\begin{equation}
T_E = T_0 + T_1 \cos \phi,
\end{equation}
which corresponds to the environmental temperature.
For $T_1 > g/C_P$ the system is convectively unstable, while for $T_1 < T_0$, the 
background temperature can never be negative.
We represent the temperature by 
\begin{equation}
T = T_0 + T_2 \cos \phi + T_3 \sin \phi. \label{Tdecomp}
\end{equation}

\noindent{\bf Viscous damping.}\label{viscous}
The viscous damping term 
may be approximated by $\nu {\bf  \nabla^{2} u} \rightarrow - \Gamma  u  = -\nu (2/\ell)^2 u$. 
Here the constant $\Gamma$ is the inverse of the viscous dissipation time scale $\tau_{vis}$.

The buoyant acceleration in the vertical direction is 
\begin{equation}
 B_z = -{\bf  g \cdot} {\bf (\delta \rho / \rho)} \ \rightarrow g \beta_T  (2T_3 /T_0) \sin^2 \phi. \label{buoy}
\end{equation}
Only a temperature (buoyancy) difference in the horizontal direction ($\phi = \pi/2$) 
gives a net torque to turn the convective roll.  

With damping due to a linear viscosity term, 
\begin{equation}
    d u/dt =  - \Gamma u
    + {g \beta_T \over T_{0}} (T-T_{E}) \sin \phi.
    \label{eqb1b}
\end{equation}

Integrating Eq.~\ref{eqb1b} over a complete cycle in $0 \le \phi \le 2 
\pi,$ gives\footnote{The two 
terms constant in space ($\phi$) generate a factor of $2\pi$ while the
integral over $\sin^2 \phi$ gives a factor of $\pi$.}
\begin{equation}
    du/dt = -\Gamma u + \Big[  {g\beta_T \over 2 T_{0} } \Big ] T_{3}
    \label{eq1c}
\end{equation}
The temperature terms in $\cos \phi$ integrate to zero because of
the $\sin \phi$ factor in the buoyancy term. Thus we have just two
terms in Eq.~\ref{eq1c}, a sink from the viscous damping and a source from 
buoyancy.

In the turbulent case, we might identify this rate of dissipation as an integral over 
the turbulent region, so it is transformed into a global quantity, 
$\varepsilon_K/\rho = u^3/\ell$.
This is the deceleration times the velocity, giving a deceleration of
$-u \vert u \vert/\ell$, so that
$\nu {\bf  \nabla^{2} u} \rightarrow - u\vert u \vert /\ell$.
The absolute value $\vert u \vert$ is used here because the characteristic time
is $\tau_{vis} = \ell/\vert u \vert $, and the deceleration is thus $-u/\tau_{vis}$.
The length scale is the depth of the convective shell (i.e., $ \ell $).
The convective speed $u$ is constant in space, so we may use a
nonlinear damping term, as implied by \cite{kolmg41}, 
\begin{equation}
    du/dt = - u\vert u \vert /\ell + \Big[  {g\beta_T \over 2 T_{0} } \Big ] T_{3}.
    \label{eq1ca}
\end{equation}

\noindent{\bf Energy conservation.\label{AEcons}}
At constant pressure, the first law of thermodynamics simplifies to
\begin{equation}
dW/dt = C_P dT/dt = \epsilon - {1 \over \rho}{\bf \nabla \cdot F} + {\varepsilon_K \over \rho },
\label{dwdt}
\end{equation}
where $\epsilon$ is the net heating-cooling rate from nuclear burning and 
neutrino emission, $C_P$ the specific heat at constant pressure, $T$ the
temperature, $\rho$ the mass (nucleon) density,  $\bf F$ the flux of radiative energy,
and $\varepsilon_K$ the volumetric heating by turbulence.
Ignoring $\epsilon$ and $\varepsilon_K$, 
\begin{equation}
\partial T /\partial t + {\bf u \cdot \nabla T} = -( {1 \over \rho}{\bf \nabla \cdot F} )/C_P.
\end{equation}
The divergence of flux of radiative luminosity is
\begin{eqnarray}
{1 \over \rho}{\bf \nabla \cdot F} &=& - {1 \over \rho}{\bf \nabla \cdot} [ -(c/3\rho \kappa) \nabla a T^4] ,
\end{eqnarray}
so that 
 \begin{eqnarray}
    \partial T/ \partial t + {\bf u \cdot \nabla}T = 
     \nu_T  \nabla^{2}T,   \label{eqb2}
\end{eqnarray}
where $\nu_T = (4acT^3/3 \rho \kappa ) / ( \rho C_P)$ is the thermal conductivity for
radiative diffusion, for constant $\kappa$ and $C_P$.

The background temperature $T_E$ is chosen such that $\nabla^2 T_E = 0$.
The radiative diffusion term is
\begin{eqnarray}
\nu_T \nabla^2 T &\approx& K(T_E-T),
\end{eqnarray}
where $K(T_E-T)$ is the extra radiative diffusion due to the deviation of temperature from  $T_E$,  and $K = \nu_T (2/\ell)^2$ is the inverse of the radiation diffusion time scale $\tau_{rad}$.

\noindent{\bf The classical Lorenz equations.\label{AclassicL}}
Following \cite{lorenz}, we correct for an aspect ratio $a$
by a factor $b = 4/(1+a^2)$ which deals with the excess in vertical over horizontal
heat conduction ($bK$ versus $K$).
Substituting for $T$ and $T_E$ in Equation~\ref{eqb2}, we have 
\begin{eqnarray}
{d T_2 \over d t} \cos \phi + {dT_3 \over dt}\sin \phi \nonumber \\
- {2uT_2  \over \ell}\sin \phi  
+  {2u T_3 \over \ell}\cos \phi   \nonumber \\
= bK(T_1 - T_2)\cos \phi - KT_3 \sin \phi.
\end{eqnarray}
Coefficients of orthogonal functions must separately sum to zero, so
\begin{eqnarray}
dT_3/dt - 2u T_2/\ell &=& -KT_3 \\
dT_2/dt + 2u T_3/\ell &=& bK(T_1 - T_2).
\end{eqnarray}
We introduce a potential temperature
\begin{equation}
T_4 = T_1 - T_2,
\end{equation}
and eliminate $T_2$. If $T_1 = g\ell/2C_P$ is independent of time,
\begin{eqnarray}
dT_3/dt =& -KT_3 + 2uT_1/\ell -2uT_4/\ell \label{eqT3} \\
dT_4/dt =& - bKT_4 + 2uT_3/\ell. \label{eqT4}
\end{eqnarray}
We define dimensionless variables
\begin{eqnarray}
\tau &=& t\ K \nonumber\\
X&=&u\ (2 / \ell K)\nonumber\\
Y&=&T_3\ (g \beta_T/\ell \Gamma K  T_0)\nonumber\\
Z&=&T_4\ (g \beta_T/\ell \Gamma K T_0), \label{nondim}
\end{eqnarray}
and use Equations~\ref{eq1c}, \ref{eqT3}, and \ref{eqT4}
to get the Lorenz equations in their usual form:
Equations~\ref{lor1}, \ref{lor2}, and \ref{lor3}.
\begin{eqnarray}
dX/d\tau =& -\sigma X + \sigma Y    \label{lor1a}\\
dY/d\tau =& -XZ +rX -Y \label{lor2a}\\
dZ/d\tau =& XY - bZ, \label{lor3a}
\end{eqnarray}
where $\tau$ is a time in thermal diffusion units, 
$\sigma =  \Gamma/K$ is the 
effective Prandtl number, and
$r= (g \beta_T  T_1/\ell \Gamma K T_0) = (g \beta_T  T_1 / \ell \Gamma^2 T_0)\sigma$ 
is the ratio of the Rayleigh number to its critical value for onset of
convection. 
The Prandtl number is the ratio of coefficients of the viscous dissipation term 
to the thermal mixing term, $\sigma = \nu/\nu_T = \Gamma/K = \tau_{rad}/\tau_{vis}$,
where $\tau_{vis} =1/\Gamma$ is the viscous damping time.

\section{APPENDIX: What Should The Prandtl Number Be?\label{prandtl}}

The original Lorenz parameters appear to be a fair choice
to represent the flow in 3D simulations (see Figure~\ref{fig1}).
This is consistent with an effective Prandtl number for the numerical simulation
of $\sigma_t \approx 10$; however, the severe reduction in degrees of freedom
from the simulations to the Lorenz model warns
against taking the numerical value too literally. We note for comparison
that water has $\sigma \approx 6$ and air has $\sigma \approx 0.7$.
Apparently Lorenz felt he was lucky (\cite{lorenzchaos}, p. 137); he took a suggested
value ($\sigma=10$) which gave chaotic behavior instead of a lower value,
actually appropriate for air, for which his equations give stable rolls.
As a mathematical example this quantitative difference in the Prandtl number
is not significant, but for physical applications, it is.

\subsection{The Microscopic Value}
Using the ratio of microscopic thermal diffusion to viscous time scale,
\cite{hansen} (page 178 and 185) suggest that $\sigma \approx 10^{-8}$. 
\cite{lorenz} finds the critical value for
$r$ for instability of steady convection to be
$r_c = \sigma(\sigma + b + 3)/(\sigma -b -1)$. For $\sigma < b+1 = 2.66\overline{6}$, 
steady convection
is always stable, so that the \cite{hansen} value would never give turbulence.
If $\sigma > b+1$, steady convection is unstable for sufficiently
large Rayleigh numbers. This precise value for instability is a characteristic of the canonical Lorenz model, and  is affected by the particular choice of dissipation function (see Appendix A).

\subsection{The Simulation Value}
In the numerical simulations, the effective Prandtl number is dominated by
the turbulent cascade, which gives mixing of both momentum and heat
at a rate determined by the largest eddy size. Thus, in numerical simulations,
the turbulent flow defines its own effective value of this parameter 
$\sigma \rightarrow \sigma_t$, and whatever value $\sigma_t$ has, it is clearly above
the threshold for instability for the system we have numerically simulated. 
Numerical simulations and experiment suggest, for developed turbulence
at high Reynolds numbers, $\sigma_t \approx 0.7$, a value typical of many common
gases.

We note that in the ``Reynolds analogy" (\cite{monin}, p. 341), Osborne Reynolds argued that the
mechanisms for transport of heat and momentum were essentially the same in
a turbulent medium, so that the effective turbulent Prandtl number should be 
of order unity ($\sigma_t  \approx 1$).

\subsection{A Cascade Estimate}\label{cascade}
Suppose we think of the Prandtl number as the relative strength of the process
which makes the velocity field isotropic to that  which converts the kinetic energy
into heat. These occur at different ends of the turbulent cascade. To better understand
what this might mean, consider $\sigma = $(time to change direction)/(time to heat).
We approximate the time to change direction by the time to halve the kinetic energy,
($ \ell/2v_{rms}$). At any length scale $\lambda$ the turbulent cascade has a
transfer rate for kinetic energy of $v_\lambda^3/\lambda = v^3/\ell$. This implies
a time spent at each length scale of $\tau_\lambda = \lambda/v_\lambda $. If each
level in the cascade is smaller on average in length scale 
by a factor $\lambda(n+1)/\lambda(n) = f$,
the total time for the cascade is $ (\ell/v)( 1 + f^{2/3} + \cdots) =(\ell/v)/(1-f^{2/3})$,
where the geometric series has been used for the summation (see also \cite{frisch}, \S7.8).
This gives an effective Prandtl number  $\sigma_t \approx 2/(1-f^{2/3})$. 
Some representative values:
for $ f = 1/e$, this gives $\sigma \approx 4.1$, for $f=1/2$ this gives $\sigma \approx 5$,
and $f=1/\sqrt{2}$ gives $\sigma \approx 10$. This argument may
tend to overestimate the Prandtl number, so that $\sigma_t$ would be smaller 
than the historical value chosen by Lorenz ($\sigma=10$).

\subsection{Renormalization Group}
The separation in size of the large scale eddies with those at the dissipation scale
suggests that this coupling might be approximated in some ingenious way.
The microscopic viscosity might be considered a ``bare" value, to be renormalized
to a ``dressed" value, in analogy to the field-theoretic treatment of interacting 
particles. \cite{leo} proposed the idea of reducing the size of a system a step at 
a time by grouping neighboring entities (in this case molecules) and treating each 
group as a single interaction. \cite{kwilson} has successfully implemented the
general idea of coarse-graining, or ``weeding out the small scales".  These general
ideas may be applied to the turbulent cascade.
\cite{yakhot} use renormalization-group (RG) analysis of turbulence with some success,
and in particular, estimate the effective Prandtl number for turbulent flow to be
$\sigma_t = 0.7179$; see also \cite{kraichnan}.  

However there is some debate:
in their review, \cite{sw98} warn ``\dots the RG method \dots
leads to suggestive results when applied to turbulence\dots However, its
application to turbulence cannot yet be called a major success, owing to the
uncontrolled approximations currently required to implement it."
A similar sentiment is found in \cite{mccomb}, p. 290.

\subsection{A Perspective}
We have captured the behavior of the pulses in the simulations, by
a Lorenz model using the parameters that Lorenz used. In this sense, these
values are relevant to our problem, although it is unclear from first
principles what the precise value of the effective turbulent Prandtl number should be.
The form for dissipation that Lorenz used is not the same as implied
by Kolmogorov, so that the threshold for instability changes \citep{am11c}.
We may interpret  the Prandtl number $\sigma$ and the Rayleigh 
number $r$  in terms of a renormalization in which
the existence of turbulence implies effective Prandtl and Rayleigh numbers for the
convective cell. The actual values of the microscopic viscosity and thermal conductivity
have little feedback on the behavior of the largest eddies; see Section~\ref{ILES}.
Magnetic fields in real stars may affect the effective Prandtl number further, which
in the classical Lorenz model affects in turn the development of instability in the 
convective roll. 

For problems which are insensitive to the details of the small scale flow, the
values of the microscopic Prandtl do not matter; the turbulent system bootstraps
to an effective dissipation for which the  rules of Kolmogorov  hold. This approximation applies to
hydrostatic (and mildly dynamic) stellar evolution, in which the burning times are long compared to sound travel times.  For these problems, ILES is appropriate. 
There are notable exceptions, such as a flame front in a medium of unmixed fuel 
and ash (SNIa  progenitor models),  for which the small scales
are important, and direct numerical simulations (DNS) of the small scales is necessary.
 
 Finally, we recall how the Lorenz model was constructed: a low order mode was chosen, a Rayleigh number was chosen just above the onset of instability  for that
 mode, and a Prandtl number was chosen which gave interesting behavior. While different parameter choices are mathematically interesting \citep{sparrow}, their physical relevance must be re-evaluated. The Lorenz equations are a spartan subset of the fluid equations which contain the germ of chaos; it is probably better
 to use them as a guide rather than a gospel.

\clearpage

\clearpage

\end{document}